\documentclass[pra,aps,10pt, twocolumn, floatfix]{revtex4-1} 

\usepackage {graphicx} 
\usepackage {amssymb} 
\usepackage {amsmath}

\usepackage{amsmath,amssymb,graphicx}

\begin{document}

\title{Generation of unipolar pulses in a circular Raman-active medium excited by few-cycle optical pulses}

\author{R. M. Arkhipov$^{1,2}$, A.V. Pakhomov$^{3,4}$, I. Babushkin$^{5,6}$, M. V. Arkhipov$^2$,   Yu.A. Tolmachev$^2$, N.N. Rosanov{$^{1,7}$}}
\affiliation{$^1$ ITMO University, Kronverkskiy prospekt 49, St. Petersburg 197101, Russia
 \\
  $^2$ Faculty of Physics, St. Petersburg State University, Ulyanovskaya 3, Petrodvoretz, St. Petersburg 198504, Russia \\
  $^3$ Department of Physics, Samara University, Moskovskoye Shosse 34, Samara 443086, Russia\\
$^4$ Department of Theoretical Physics, Lebedev Physical Institute, Novo-Sadovaya Str. 221, Samara 443011, Russia\\
  $^5$ Institute of Quantum Optics, Leibniz University Hannover,
  Welfengarten 130167, Hannover, Germany \\
$^6$ Max Born Institute, Max Born Str. 2a, Berlin, Germany\\
$^7$ Vavilov State Optical Institute,
199053, Russia, Saint-Petersburg, Kadetskaya liniya V.O., 14/2
}

\date{\today}

\begin{abstract}
 We study theoretically a new possibility of unipolar pulses
  generation in Raman-active medium excited by a series of few-cycle 
optical pulses. We consider the case when the Raman-active particles are uniformly
  distributed along the circle, and demonstrate a 
  possibility to
  obtain a unipolar rectangular video pulses with an arbitrarily long
  duration, ranging from a minimum value equal to the natural period
  of the low frequency vibrations in the Raman-active medium.
\end{abstract}


\maketitle

\section{Introduction}

Generation of the extremely-short optical pulses of the
 single and half optical cycle duration
is an area of principal interest in modern optics because of the
rapidly growing fields of their applications
\cite{dudley06,skryabin10, leblond2013, Gallmann, Chini,Landsman,Strelkov}. For example, they can be used to study the dynamics of the individual motion of electrons in atoms and molecules, high resolution control of ultrafast processes in matter, nonlinear optics and medicine. Such extremely-short optical pulses contain broad-band spectrum up to zero frequencies. 
In contrast to conventional optical pulses
  which are bipolar, that is, 
$\int\limits_{-\infty}^{+\infty} E(t)dt=0$,
  the half-cycle pulses can contain
a DC component of the electric field,
that is, $\int\limits_{-\infty}^{+\infty} E(t)dt\ne0$, and
often referred to as unipolar
pulses. Possession of this characteristic property makes unipolar pulses 
ideally suited for the control of charges dynamics in matter.
Half-cycle pulses have been obtained experimentally
in terahertz range \cite{reimann07,You,Gao} and have been used for the ionization and to produce novel dynamic states in Rydberg atoms \cite{Gao, Jones,Bensky,Wetzels}. Unipolar half-cycle pulses can be produced experimentally when irradiating a double foil target with intense few-cycle laser pulses \cite{Wu}. 

Possibility of unipolar pulse generation was studied theoretically by
different authors. Such pulses can be obtained for instance when
an initially bipolar ultra-short pulse propagates in a
nonlinear resonant medium
\cite{Leblond,vysotina2006extremely,rosanov2008few, kazantseva2001propagation, kozlov2011generation, Kalosha1999, song2010unipolar, song2015origin}, or, alternatively, in Raman-active medium (RAM) in the regime of self-induced transparency \cite{belenov1992dynamics, belenov1994propagation}.
In recent years,  interest to sources of such unipolar pulses has been closely related to the study of new ways of generation of high power attosecond pulses \cite{orlando2009generation, pan2013generation, feng2015unipolar}. 

Interaction of electromagnetic waves with ensembles of nanoparticles have 
attracted considerable research interest over the last years. Recent progress in the development of the plasmonic nanoparticles fabrication techniques \cite{Kreibig1995, Feldheim2001} led to the emergence of their various applications as optical waveguides \cite{Maier2003, Quinten1998}, surface enhanced Raman scattering processes \cite{Moskovits1985, Bachelier2004}, 
high quality optical resonators \cite{Alu2006, Citrin2005, Burin}, antennas and detectors \cite{King} etc. The various types of particles arrangements - linear \cite{Maier2001, Maier2002, Maier2004, Gozman2008, arkhipov2014transient} and circular \cite{Fikioris2011, Fung2008, Burin, Citrin2005, Polishchuk2009, arkhipov2015transient} have been studied for these purposes. 
Among others, circular arrays of nanoparticles have attracted great attention for a number of reasons. In particular, circular arrays of microsize particles can possess bound whispering gallery modes having an extremely high quality factor \cite{Burin, King}. A circular arrays of oscillators may exhibit a resonant dipole collective response and thus serve as electric and magnetic resonators at optical frequencies \cite{Alu2006}. Ring-shaped geometry of nanoparticles array was also found to be the optimal for engineering the highly radiative modes with suppressed radiative losses \cite{Citrin2005}. 
	 
In a prior paper \cite{arkhipov2016}, we have predicted theoretically
a novel possibility of unipolar pulse generation in the case when a linear string 
of Raman-active particles is excited by a train of few-cycle 
optical pulses propagating along the medium at the velocity
faster than the velocity of light in vacuum $c$. Such superluminal
excitation occurs when an optical pulse is incident on a 
flat screen at some angle $\beta$. In this case, cross-section point of
pulse and medium propagates along the medium at the velocity
$V=c/\sin\beta >c$ \cite{bolotovskii1972,ginzburg1979}. Similar situation of superluminal excitation of a resonant medium composed of classical harmonic oscillators was studied in \cite{arkhipov2014transient,arkhipov2015transient}. 
However, the method proposed in \cite{arkhipov2016} seems to have a 
number of limitations. First of all, pulse duration is strongly limited by the finite length of linear string.
 Next, increasing the pulse duration leads to a significant decrease of the pulse amplitude. 

In this paper, we consider an array of Raman-active particles which has
circular or helix form, in contrast to a linear geometry considered 
before. We demonstrate, that such geometry also allows generation of unipolar pulses. We show
also, that it provides a number of significant advantages
over the linear one. Primarily, the circular geometry allows to achieve an arbitrary pulse 
duration, in contrast to the linear case. Besides, the amplitude of the 
long unipolar pulses can be significantly increased. In general, 
circular geometry allows more efficient control of the pulse duration, 
amplitude and shape.

\section{Physical consideration and theoretical model}

Let us consider the physical picture of unipolar pulse generation in RAM \cite{arkhipov2016}. This method is similar to well-known method of optical control over elementary molecular motion  with sequences of 
femtosecond pulses \cite{weiner1990, weiner1991}. Two few-cycle pulses
separated by the time interval $T_p=T_0/2$ equal to half-period (or
odd number of half-periods) $T_0/2$ of natural vibrations of low
frequency oscillator in RAM are launched in RAM. In this case, the
first pulse causes the motion of the low frequency oscillator and
second pulse can stop this motion.

The interaction of ultra-short pulse with RAM can be described in terms
of classical theory by using a model of nonlinearly bonded oscillators: high frequency oscillator (HFO), for example electron in molecule, and low frequency oscillator (LFO), for example, nucleus in molecule~\cite{akhmanov1997physical}. Let us denote the normal vibration coordinate of LFO as $y$ and the normal vibration coordinate of HFO as $x$. The potential energy of this system can be written as \cite{akhmanov1997physical} 

\begin{equation}
U(x,y)=\frac{1}{2}\alpha x^2 + \frac{1}{2}\beta y^2 + \frac{1}{2}\gamma x^2y,
\label{eq:U} 
\end{equation} 
where $\alpha$ and $\beta$ denote elastic constants of intramolecular
bonds, term $\gamma$ describes the non-linear interaction
between HFO and LFO. In accordance with Eq.~(\ref{eq:U}), the
equations describing the interaction of LFO and HFO with the electric
field of the pump pulse take the form \cite{akhmanov1997physical}:

\begin{eqnarray}
\ddot{x}+\Gamma_e \dot{x} + \Omega_{0}^2 x=\frac{e}{m}E - \frac{\gamma}{m}xy,
\label{eqx} \\
\ddot{y}+\Gamma_n \dot{y} + \omega_{0}^2 y=-\frac{\gamma}{2M}x^2.
\label{eqy}
\end{eqnarray}

 Here, $M$ and $m$ are reduced masses of the LFO and HFO, $E$ denotes the electric field, $\Gamma_e$ and $\Gamma_n$ are the damping rates of HFO and LFO respectively,  
  $\omega_0=2\pi/T_0=\sqrt{\frac{\beta}{M}}$ is the resonance frequency of LFO, $\Omega_0=2\pi/T=\sqrt{\frac{\alpha}{m}}$ is the natural frequency of HFO. 
  It is seen from ~\eqref{eqy} that LFO displacement $y$ is
  proportional to $x^2$ and hence proportional to $E^2(t)$. As a result,
   the spectrum of the generated field can contain the constant component.

\begin{figure}[htpb]
\includegraphics[width=1\linewidth]{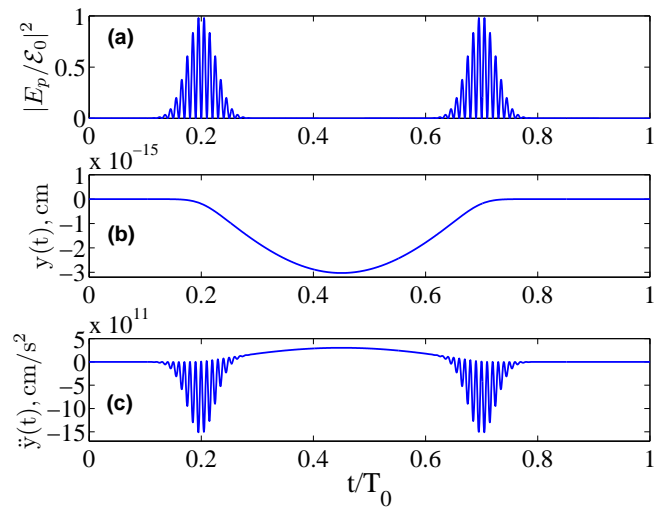}
\caption{ Time dependence of $|E_p/\mathcal{E}_0|^2$ of the excitation few-cycle pulses (a), LFO displacement $y(t)$ (b), and LFO acceleration $\ddot{y}(t)$ (c). Parameters: $\Omega_0=10^{15}$ rad/s, $\omega_0=10^{13}$ rad/s ($T_0=2\pi/\omega_0=0.628 ps$), $e=-4.8\cdot 10^{-10}$ ESU, $m=9.1\cdot10^{-28}$ g (electron mass), $m=1.6\cdot10^{-24}$ g (proton mass), $\gamma=1000$, $\omega=0.5\Omega_0$ ($T=2\pi/\omega$), $\tau=5T$, $T_p=T_0/2$, $E_0=10^5$ ESU, $\Gamma_e=\Gamma_n=0$.}
 \label{fig0}
\end{figure}

Let us assume, that the medium is excited by two Gaussian pulses
\begin{gather}
\nonumber
E_p =\mathcal{E}_0 \exp\left[-t^2/\tau^2\right]\sin\omega t
 \\ +\mathcal{E}_0 
\exp\left[-\left(t-T_p\right)^2/\tau^2\right]\sin\omega(t - T_p),
  \label{eq:Epump}
\end{gather}
 separated by the time interval $T_p=T_0/2$. Here, $\mathcal{E}_0 $ is pulse amplitude, $\omega$ is the pulse carrier frequency, 
$\tau$ is pulse duration. To effectively excite LFO oscillator we suppose $\omega\gg \omega_0$ and $\omega_0\tau\ll 1$.
 The results of the numerical integration 
 of the set of the equations ~(\ref{eqx})-(\ref{eqy}) with the
 electric field ~(\ref{eq:Epump}) are illustrated in
 Fig.~\ref{fig0}. It is seen that the first pulse initiates the motion
 of LOF and the second pulse stops it, see Fig.~\ref{fig0}b. Thus
 polarization of LFO has the shape of a half-sine wave with
 the duration $T_0/2$, that does not
 change its sign (see Fig.~\ref{fig0}b). Electric field radiated by RAM
 far away from the medium is proportional to the second derivative of
 the polarization (charge acceleration), which is plotted in
 Fig.~\ref{fig0}c. It is seen that this acceleration contains high
 frequency oscillations at the moments of the start and
 stop of the HFO motion, and has the shape of half wave between them. For the obtained pulse to be unipolar the only central half-sine wave should be kept, while the high frequency components distorting the pulse shape should be someways suppressed. These high frequency oscillations can be cut off by spectral filter and we neglect them in further consideration.
  The media which can support Raman term can be very
  different in their nature, including molecular gases
  \cite{Nibler}, solids \cite{schrader95} or
  semiconductors \cite{scott69,klyukanov01,klyukanov02}. 

 Thus we approximate the half-wave of polarization in Fig.~\ref{fig0}b by sine function, and
  the expression for the polarization of the medium is
 proportional to~\cite{akhmanov1997physical}:
\begin{equation}
  \label{eq:2}
  P(t) \sim \sum_{k=0}^{N_{p}-1} e^{-\gamma t'} \sin\left[\omega_0 t' \right] \Theta \left[t' \right].
\end{equation}
Here $t' = t - k T_p$, $\Theta(t')$ is Heaviside step function,
$\gamma$ is the damping rate of oscillator, $T_p$ is the pulse
repetition period, $N_{p}$ is the number of pulses.

\section{Circular string of Raman-active particles}

\begin{figure}[htpb]
\includegraphics[width=1\linewidth]{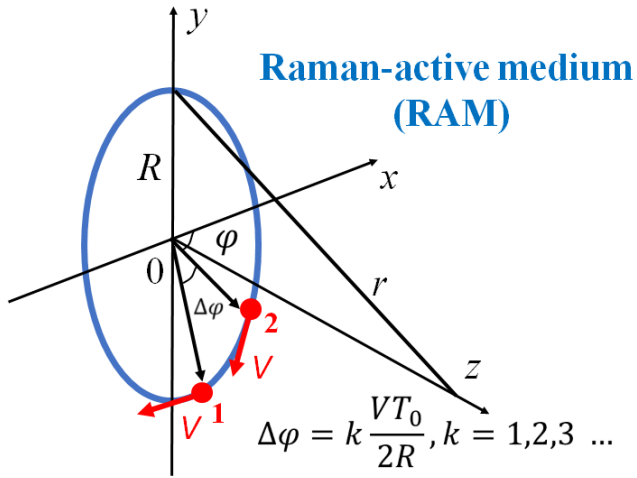}
\caption{Circular string of RAM (blue circle) is excited by two extremely-short light pulses 1 and
 2 (red circles) propagating with velocity $V$ along the circle. The
 angular distance between spots is $\Delta\varphi$. Radiation from
 the medium is observed in a point on $z$ axis far away from the circle, the distance between
 the circle and the point of observation is $r$.}
 \label{fig1}
\end{figure}

Figure ~\ref{fig1} shows the RAM, located on the string in a form
of a circle of radius $R$. The linear
density of oscillators is assumed to be constant. Over the
round-trip, two spots of light 1
and 2 of extremely-short pulses are moving at a constant velocity $V$
and excite the Raman-active particles. Note that the speed of
excitation spot on the circle can be greater than the speed of light
in vacuum $c$, see reviews \mbox{\cite{bolotovskii1972,
  ginzburg1979}}. To realize such motion along the circles one can suggest to use laser beam scanners (laser beam deflectors), i.e. the devices which are used to deflect the laser beam \cite{Romer, Arkhipovcamera}. Note that
any points located on the $Oz$ axis (perpendicular to
the circle plane) are equally distant from the points of
the RAM. 

The electric field generated by the dipoles after a
short and rapid rise will remain constant. Therefore, the rectangular unipolar pulse with a duration equal to the duration of excitation will be formed at the observation point. 
 We assume that the angular distance between the excitation regions 1 and 2 is $\Delta\varphi = VT_0/2R$, which provides excitation of one half-cycle of oscillation of RAM. It is also assumed, that the oscillator damping can be neglected. 
Electric field on $z$ axis at a large distance from the circle is proportional to the oscillator acceleration, i.e. second time derivative of the dipole displacement plotted in Fig.~\ref{fig0}c. This acceleration contains a high frequency component arising at the moments of start and stop of the dipole vibrations as well as the ``half-wave'' part of low frequency oscillation in RAM, $\omega_0=2\pi/T_0$. 
Assuming the high-frequency components to be filtered out, we get the emitted electric field in the form of the half-sine pulse.
Therefore, the total electric field on $z$ axis is given as:
\begin{equation}
E(t) =E_0 \sum_{k=0}^{1}  \int\limits_0^{2\pi N}  \sin\Big[\omega_0 f_{\varphi} \Big] \Theta\Big[f_{\varphi} \Big] d\varphi.
\label{eqEphi}
\end{equation}
Here $f_{\varphi}\equiv t-\frac{R\varphi}{V}-\frac{r}{c}-(k-1)T_p$,
$N$ - total number of rotations of excitation points over the circle,
$T_p =R\Delta\varphi/V$  - time interval between exciting pulses,
$r/c$ – light propagation time from the circle to the observer,
$V$ – velocity of excitation, $E_0$ is some constant.   

\begin{figure}[htpb]
\includegraphics[width=1\linewidth]{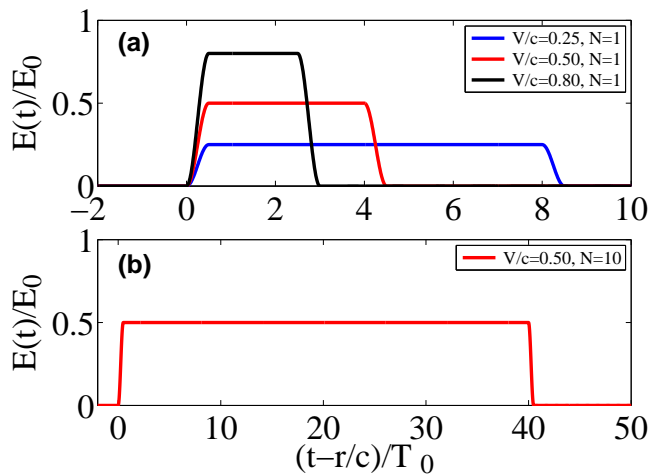}
\caption{Results of the numerical solution of the integral
  \eqref{eqEphi} for $\frac{\omega_0 R}{c}=2$ and (a) $N=1$ and different values of the velocity of excitations
  $V$; (b) $N=10$ and $V/c=0.5$.}
 \label{fig3}
\end{figure}

The results of calculation of Eq.~(\ref{eqEphi}) are shown in
Fig.~\ref{fig3} for different values of $V/c$ and $N=1$ (a) and total
number of rotations $N=10$ if $V/c=0.5$ (b). As can be seen from the Fig.~\ref{fig3}a, the system produces a rectangular unipolar pulse whose duration is equal to the entire duration of the excitation process. This is because the elementary excitations (half-waves) coming from different parts of the circle at different times are summed. Pulse duration can be thus easily controlled by the 
total number of rotations of excitation pulses over the string.
The pulse amplitude is given by (see \eqref{eqA2} in Appendix)
\begin{eqnarray}
A = \frac{2E_0 V}{\omega_0 R}.
\end{eqnarray}
 It can be seen from Fig.~\ref{fig3}a that the unipolar pulse
duration increases with the increase of $\Delta\varphi$, i.e. with the
increase of excitation velocity $V$. Increasing the number of
rotations $N$ 
leads to an increase of the
pulse duration whereas the amplitude remains the same, which is illustrated in Fig.~\ref{fig3}b for $N = 10$. 
This effect can not be achieved with the previously studied linear arrangement of particles \cite{arkhipov2016}, since
the pulse amplitude and duration vary inversely.

When $\Delta\varphi=2\pi$, an excitation of RAM by
a single spot of light takes place. In this case, to satisfy
the condition of half-wave excitation the velocity of excitation
should be significantly larger. To realize this rapid rotation ultrafast laser beam deflectors can be used. This in principle makes some
limitations on the method proposed in this section. Hence, it is easier to use two excitation spots instead of one.

\section{Raman active particles are distributed along the cylindrical
  helix}

\begin{figure}[htpb]
\includegraphics[width=1\linewidth]{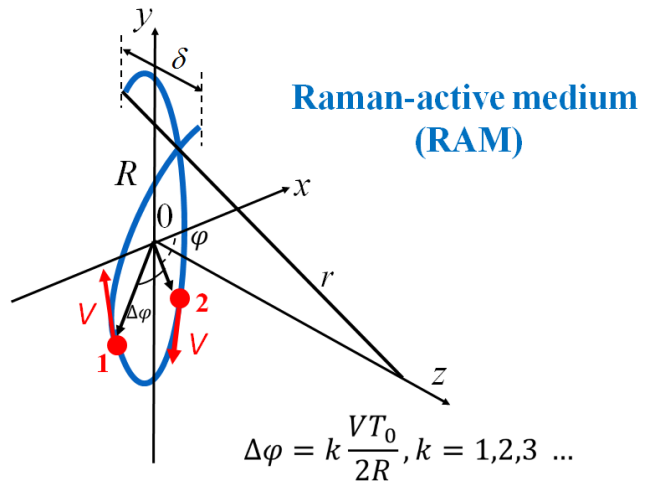}
\caption{Raman-active particles are distributed along the cylindrical helix
of the pitch distance $\delta$
and excited by two extremely-short light pulses 1 and 2 (red circles)
propagating with velocity $V$ along the helix. The
angular distance between spots is $\Delta\varphi$. The medium radiation
 is observed in a point on $z$ axis far away from the helix, 
the distance between the helix starting point and the observation point is $r$.}
\label{fig4}
\end{figure}

The rectangular unipolar pulses will also arise if one
arranges the delay in the emitted field arrival to the
observation point in such a way that this delay varies
linearly with increasing of the polar angle $\varphi$ on the
roundtrip. Such delay will take place if we cut the
circle at some point and convert it to one
turn of a cylindrical helix (see Fig.~\ref{fig4}).
It is interesting to note that helicoidal structures are widely 
spread in natural systems, ranging from DNA, RNA and other
molecules to sea shells and even spiral galaxies \cite{Cross}.

We take the observation point far enough from the helix, i.e. the distance to the 
helix starting point $r\gg R,\delta$, or, alternatively, in the focal plane of the focusing lens,
orthogonal to the $Oz$ axis and spaced at the distance $r$. Oscillators are still excited by 
two extremely-short light pulses propagating with velocity $V$ along the helix with the $T_0/2$ delay. 
The equation of such cylindrical helix has the form
\begin{eqnarray}
x=R\cos\varphi, y=R\sin\varphi, z=\delta\varphi/2\pi, 
\label{eqhelix}
\end{eqnarray}
 here $\delta$ is the helix pitch distance, see Fig.~\ref{fig4}.

\begin{figure}[htpb]
\includegraphics[width=1\linewidth]{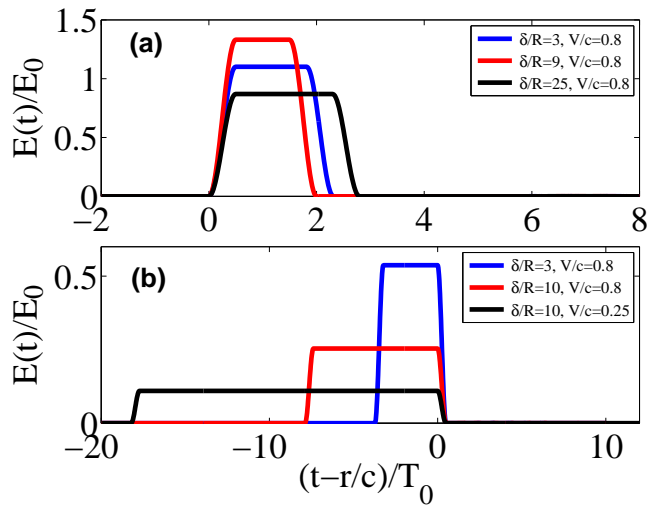}
\caption{Results of the numerical solution of the integral
  \eqref{eqEphi1} for $\frac{\omega_0 R}{c}=2$, when the exciting extremely-short pulses propagate along the helix
 (a): in the clockwise direction; (b): in the anticlockwise direction.}
 \label{fig5}
\end{figure}

Unipolar pulse duration in this case will be governed not only by the excitation velocity $V$, but also by 
the helix pitch distance $\delta$. Indeed, the effective contribution for an arbitrary oscillator in the spiral to the 
resulting emission will be determined by both its emission onset delay due to exciting velocity $V$ being finite and
by its distance from the observation point with respect to other oscillators, as this distance varies along the helix 
in contrast to the previously studied circular configuration. 
Depending on the direction of helix traversal by the exciting pulses - clockwise or anticlockwise as viewed from the observation point -
these factors can contribute either jointly or contrarily to the generated unipolar pulse parametes. 
Mathematical expression determining the pulse shape has the form:
\begin{eqnarray}
E(t) =E_0 \sum_{k=0}^{1}  \int\limits_0^{2\pi S}  \sin\Big[\omega_0 g_{\varphi} \Big] \Theta\Big[g_{\varphi} \Big]d\varphi.
\label{eqEphi1}
\end{eqnarray}
Here, $g_{\varphi}\equiv
t\mp\frac{\varphi}{V}\sqrt{R^2+(\frac{\delta}{2\pi})^2}-\frac{1}{c}(r-\frac{\delta\varphi}{2\pi})-(k-1)T_p$,
$S$ - number of helix coils. The first delay term in $g_{\varphi}$ stands for the emission onset delay and its sign is determined by the
direction of helix traversal, while the second delay term denotes the emission arrival variation to the observation point, arising
from the helix extension along the $Oz$ axis.

	The results of numerical calculation of the integral
        \eqref{eqEphi1} for different values of $\delta$ and $V$ are
        plotted in Fig.~\ref{fig5} with $S=1$ and for different directions of helix traversal. It stands to mention, that dependence     
        $g_{\varphi}({\varphi})$ is linear, thus leading to the rectangular pulse shape according to the findings of the previous section,
        but with the complex aggregate multiplying coefficient $\frac{\delta}{2\pi c}\mp\frac{1}{V}\sqrt{R^2+(\frac{\delta}{2\pi})^2} $, that 
         actually sets the duration and amplitude of the emitted unipolar pulse. 
         In the case of the clockwise traversal direction (sign $"-"$ in \eqref{eqEphi1}),  if $V>c$, this coefficient grows 
        monotonically with the helix pitch distance $\delta$, but when $V<c$, this coefficient reaches maximum value,
        corresponding to the minimal unipolar pulse duration, for:
\begin{eqnarray}
\delta_m=\frac{2\pi R}{\sqrt{(\frac{c}{V})^2-1}}.
\label{eqDelta}
\end{eqnarray}
        
        The latter case is illustrated by Fig.~\ref{fig5}a, with the shortest pulse for the helix pitch distance close to given by \eqref{eqDelta}.
        For the anticlockwise traversal direction (sign $"+"$ in \eqref{eqEphi1}), the coefficient at ${\varphi}$
        appears to be monotonically increasing function of the helix pitch distance $\delta$ for the fixed excitation velocity $V$,
        what enables to get unipolar pulses much longer than from the planar circle (see  Fig.~\ref{fig5}b).
        
Similarly to the circular case the pulse amplitude is given by (see Eqs. (\ref{eqA2}), (\ref{eqA35}) in Appendix): 

\begin{eqnarray}
A = \frac{2E_0}{\omega_0 |\frac{1}{V}\sqrt{R^2+(\frac{\delta}{2\pi})^2}\mp\frac{\delta}{2\pi c}|}.
\end{eqnarray}
        As the duration of the rectangular unipolar pulse increases with the proper changing of $\delta$ and $V$, its amplitude correspondingly decreases, thus
        keeping the whole pulse area constant. The total pulse duration can be, however, effectively controlled by the number of helix coils enabling to get arbitrarily long pulses.
        Leading and trailing edges of the pulse are determined by the two halves of the sinusoidal signal in Fig.~\ref{fig0}b 
       (see exact Eqs. (\ref{eqA31}), (\ref{eqA33}) in Appendix). It is         
      interesting to note that if the system is excited by the even number of pulses, generation of trapezoidal pulses becomes possible.

\section{Conclusions}
	
In conclusion, we studied the features of the
optical emission of a Raman-active medium
placed uniformly along a circle or helix and
excited by a sequence of two few-cycle optical
pulses. 
Theoretical model of RAM composed of high frequency oscillator
(electron oscillator) and low frequency (nucleus) oscillator was
applied to cover the main aspects of the RAM excitation response. 

Our theoretical analysis and numerical simulations revealed the
possibility of unipolar pulse generation of approximately
rectangular form. The geometry considered here allows
to get over the limitations imposed by the linear one, discussed earlier 
in \cite{arkhipov2016}, and
provides more efficient control of the emitted pulse parameters.
In particular, the method allows to obtain unipolar pulses of an
arbitrarily large duration governed by the total number of rotations of the 
excitation spots along the string. The lower pulse duration 
limit is determined by one half of the natural period of the low 
frequency oscillators of RAM. These unipolar pulses may be of interest 
in the development of optical switches and logic elements alternative to switches which use bistability and multistability 
\cite{gibbs2012optical,babushkin00a}.

\section*{Funding Information}

\textbf{Funding.}  This work was partially financially supported by
Government of Russian Federation, Grant 074-U01 and Russian Foundation
for Basic Research, Grant No. 16-02-00762.  I.~B. is thankful
  for the support of German Research Foundation (BA 4156/4-1) and Volkswagen
Foundation (Nieders. Vorab. ZN3061). 

\section{Appendix. Analytical solution of the equation (6)}
We provide here the exact solution of the integral Eq.~(\ref{eqEphi}) setting $N_{p}=2$, $N=1$.
To treat Eq.~(\ref{eqEphi}) analytically, Heaviside step functions $\Theta\Big[f_{\varphi} \Big]$ under the integral sign 
should be carefully expanded.
Since the function argument $f_{\varphi}$ depends on both time $t$ and polar angle $\varphi$, resulting
integral will have varying form on different stages of the excitation process.

During the initial time interval 
$\frac{r}{c}<t<\frac{r}{c}+T_p$, when the first pulse has started exciting the medium, while the second pulse still hasn't, denoting
$t''\equiv t-\frac{r}{c}$,  $\frac{1}{W} \equiv \frac{R}{V}$, one obtains:
\begin{eqnarray}
E(t) = E_0 \int_{0}^{Wt''}\sin [\omega_0 (t'' - \frac{\varphi}{W}) ]d\varphi = \\
\nonumber
= E_0 \frac{W}{\omega_0}(1-\cos [\omega_0 t'']).
 \label{eqA31}
\end{eqnarray}
The field amplitude grows from zero value at $t"=0$ to the maximum value $\frac{2E_0 W}{\omega_0}$ at $t"=T_p$.
During the next stage $\frac{r}{c}+T_p<t<\frac{r}{c}+\frac{2\pi R}{V}$ the medium is excited by both pulses and Eq.~(\ref{eqEphi}) yields:
 
\begin{eqnarray}
 \nonumber 
E(t) = E_0 (\int_{0}^{Wt''}\sin [\omega_0 (t'' - \frac{\varphi}{W}) ]d\varphi +\\
 \nonumber
 + \int_{0}^{W(t''-T_p)}\sin [\omega_0 (t'' - \frac{\varphi}{W}-T_p) ]d\varphi )= \\
= E_0 \frac{W}{\omega_0}(2-\cos [\omega_0 t'']-\cos [\omega_0 (t''-T_p)]).
 \label{eqA32}
\end{eqnarray}

It is seen, that since $T_p=T_0/2$ the previous expression can be simplified and equal to constant, which determines the peak value $A$ of the rectangular pulse:
\begin{equation}
 A = \frac{2E_0W}{\omega_0}.
 \label{eqA2}
\end{equation}

When the first pulse has passed through the medium, but the second one still hasn't: $\frac{r}{c}+\frac{2\pi R}{V}<t<\frac{r}{c}+\frac{2\pi R}{V}+T_p$, one gets:

\begin{eqnarray}
 \nonumber 
E(t) = E_0 (\int_{0}^{2\pi}\sin [\omega_0 (t'' - \frac{\varphi}{W}) ]d\varphi + \\
 \nonumber
+ \int_{0}^{W(t''-T_p)}\sin [\omega_0 (t'' - \frac{\varphi}{W}-T_p) ]d\varphi ) =\\
 \nonumber
= E_0 \frac{W}{\omega_0}(\cos[\omega_0 (t''-\frac{2\pi}{W})]-\cos [\omega_0 t'']+ \\
 \nonumber
+1-\cos [\omega_0 (t''-T_p)]) = \\ 
= E_0 \frac{W}{\omega_0}(1+\cos[\omega_0 (t''-\frac{2\pi}{W})]).
 \label{eqA33}
\end{eqnarray}
The field amplitude now decreases from the maximum value $\frac{2E_0 W}{\omega_0}$ at $t"=\frac{2\pi R}{V}$ to zero value at $t"=\frac{2\pi R}{V}+T_p$.

Finally, if $t>\frac{r}{c}+\frac{2\pi R}{V}+T_p$, after the end of the transient processes expression for the electric field has the form :
\begin{eqnarray}
 \nonumber 
E(t) = E_0 (\int_{0}^{2\pi}\sin [\omega_0 (t'' - \frac{\varphi}{W}) ]d\varphi + \\
 \nonumber
+ \int_{0}^{2\pi}\sin [\omega_0 (t'' - \frac{\varphi}{W}-T_p) ]d\varphi ) = \\
 \nonumber
= E_0 \frac{W}{\omega_0}(\cos[\omega_0 (t''-\frac{2\pi}{W})]-\cos [\omega_0 t''] + \\
+ \cos[\omega_0 (t''-\frac{2\pi}{W}-T_p)]-\cos [\omega_0 (t''-T_p)]).
 \label{eqA34}
\end{eqnarray}

Note, that if $T_p=T_{0}/2$ this expression gives zero value of the electric field $E(t)=0$.

When the helix is considered instead of the circle, described by \eqref{eqEphi1}, all the obtained expressions \eqref{eqA31}--(\ref{eqA34})
stay valid, if we substitute $1/W$ with:
\begin{eqnarray}
 \frac{1}{W}=\left|\frac{1}{V}\sqrt{R^2+(\frac{\delta}{2\pi})^2}\mp\frac{\delta}{2\pi c}\right|.
 \label{eqA35}
\end{eqnarray}





\bigskip


\end{document}